\documentstyle[prl,aps,psfig,preprint]{revtex}

\def\prl{{\it Phys. Rev. Letters}}
\def\apj{{\it Astrophys. J.}}
\def\mnras{{\it Monthly Notices Royal Astr. Soc.}}
\def\noi{\noindent}

\def\fun#1#2{\lower3.6pt\vbox{\baselineskip0pt\lineskip.9pt
  \ialign{$\mathsurround=0pt#1\hfil##\hfil$\crcr#2\crcr\sim\crcr}}}
\begin{document}

\title{THE CURIOUS ADVENTURE OF THE ULTRAHIGH ENERGY COSMIC RAYS}
   
\author{F. W. Stecker}

\address{Laboratory for High Energy Astrophysics, Code 661,
NASA/Goddard Space Flight Center, Greenbelt, MD 20771, USA.}

\maketitle

\vspace{0.8cm}

\begin{abstract}

I will discuss the mysteries involving the production and extragalactic
propagation of ultrahigh energy cosmic rays and suggested possible solutions.

\end{abstract}


\section{Introduction}

About once per century per km$^2$ of the Earth's surface, a giant shower of 
charged particles produced by a primary particle with an energy greater than 
or equal to 6 joules (100 EeV = $10^{20}$ eV) plows through the Earth's 
atmosphere. The showers which they produce can be detected by arrays of 
scintillators on the ground; they also announce their presence by producing a 
trail of ultraviolet flourescent light, exciting the nitrogen atoms in the 
atmosphere. The existence of such showers has been known for almost four
decades (Linsley 1963). The number of giant air showers detected from 
primaries of energy greater than 100 EeV has grown into the double digits and 
can be expected to grow into the hundreds as new detectors such as the 
``Auger'' array and the ``EUSO'' (Extreme Universe Space Observatory) 
and ``OWL'' (Orbiting Wide-Angle Light Collectors) satellite detectors 
come on line. These phenomena present an intriguing 
mystery from two points of view: (1) How are particles produced with such 
astounding energies, eight orders of magnitude higher than are produced by 
the best man-made terrestrial accelerators? (2) Since they are most likely 
extragalactic in origin, how do they reach us from extragalactic distances without exhibiting the predicted cutoff from interactions with the 2.7K cosmic 
background radiation? In these lectures, I will consider possible solutions 
to this double mystery.

\section{The Evidence}

Figure 1 shows the data on the ultrahigh energy cosmic ray spectrum from
the Fly's Eye and Akeno detectors. Other data from Havera Park and
Yakutsk may be found in the review by Nagano and Watson (2000) are are 
consistent with Figure 1. Additional data are now being obtained by the 
HiRes detector array and should be available in the near future (Abu-Zayyad, 
private communication).

For air showers produced by primaries of energies in the 1 to 3 EeV range, 
Hayashida, {\it et al.} (1999) have found a marked directional anisotropy 
with a 4.5$\sigma$ excess from the galactic center region, a 3.9$\sigma$ 
excess from the Cygnus region of the galaxy, and a 4.0$\sigma$
deficit from the galactic anticenter region. This is strong evidence that
EeV cosmic rays are of galactic origin. 

As shown in Figure 2, at EeV energies, the primary particles
appear to have a mixed or heavy origin, trending toward a protonic
origin in the higher energy range around 30 EeV (Bird, {\it et al.} 1993;
Abu-Zayyad, {\it et al.} 2000). This trend, together with evidence of a
flattening in the cosmic ray spectrum on the 3 to 10 EeV energy range
(Bird, {\it et al.} 1994; Takeda {\it et al.} 1998) give evidence for
a new component of cosmic rays dominating above 10 EeV energy. 

The apparent isotropy (no galactic-plane enhancement) of cosmic rays above
10 EeV ({\it e.g.} Takeda, {\it et al.} 1999), together with
the difficulty of confining protons in the galaxy at 10 to 30 EeV energies,
provide significant reasons to believe that the cosmic-ray component above 
10 EeV is extragalactic in origin. As can be seen from Figure 1, this
extragalactic component appears to extend to an energy of 300 EeV.
Extention of this spectrum to higher

\begin{figure}
\centerline{\psfig{figure=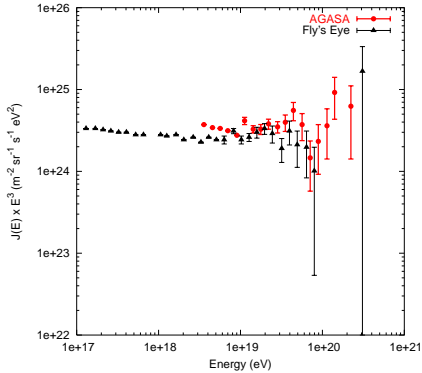,height=8cm}}

\caption{The ultrahigh energy cosmic ray spectrum data from Fly's Eye and
Akeno.}

\end{figure}

\noi energies is conceivable because such 
cosmic rays, if they exist, would be too rare to have been seen with
present detectors. We will see in the next section that the existence of 300 
EeV cosmic rays gives us a new mystery to solve.

\section{The GZK Effect}

Thirty five years ago, Penzias and Wilson (1965) reported the discovery of
the cosmic 3K thermal blackbody radiation which was produced very
early on in the history of the universe and which led to the undisputed
acceptance of the ``big bang'' theory of the origin of the universe. Much
more recently, the Cosmic Background Explorer (COBE) satellite confirmed this
discovery, showing that the cosmic background radiation (CBR) has the 
spectrum of the most perfect thermal blackbody known to man. COBE data also 
showed that this radiation (on angular scales $ > 7\deg$) was isotropic to a 
part in $10^5$ (Mather {\it et al.} 1994). The perfect thermal character 
and smoothness of the CBR proved
conclusively that this radiation is indeed cosmological and that, at the 
present time, it fills the entire universe with a 2.7K spectrum of radio to 
far-infrared photons with a density of $\sim 400$ cm$^{-3}$.

\begin{figure}
\centerline{\psfig{figure=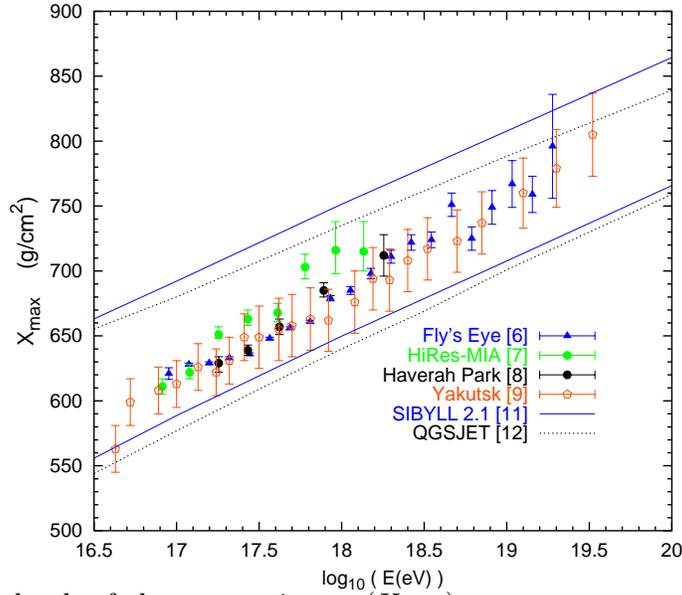,height=8cm}}

\caption{Average depth of shower maximum ($X_{max}$) {\it vs.} energy
compared to the calculated values for protons (upper curves) and Fe
primaries (lower curves) (from Gaisser 2000; see references therein).}

\end{figure}

\begin{figure}
\centerline{\psfig{figure=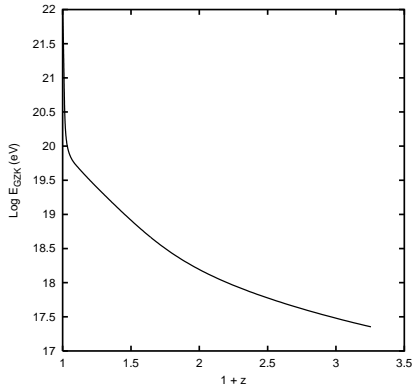,height=8cm}}

\caption{The GZK cutoff energy versus redshift (Scully and Stecker 2000).}

\end{figure}

Shortly after the discovery of the CBR, Greisen (1966) and Zatsepin and Kuz'min
(1966) predicted that pion-producing interactions of ultrahigh energy cosmic 
ray protons with CBR photons of target density $\sim$ 400 cm$^{-3}$ should 
produce a cutoff in their spectrum at energies greater than $\sim$ 50 EeV. 
This predicted effect has since become known as the GZK 
(Greisen-Zatsepin-Kuz'min) effect. Shortly after the GZK papers, Stecker (1968)
utilized data on the energy dependence of the photomeson production cross 
sections and inelasticities to calculate the mean energy loss time for protons
propagating through the CBR in intergalactic space as a function of energy.
Based on his results, Stecker (1968) then suggested that the particles of 
energy above the GZK cutoff energy (hereafter referred to as trans-GZK 
particles) must be coming from within the ``Local Supercluster'' of which we 
are a part and which is centered on the Virgo Cluster of galaxies. Thus, the 
``GZK cutoff'' is not a true cutoff, but a supression of the ultrahigh energy
cosmic ray flux owing to a limitation of the propagation distance to a few
tens of Mpc.

The actual position of the GZK cutoff can differ from the 50 EeV predicted
by Greisen. In fact, there could actually be an {\it enhancement} at or near 
this energy owing to a ``pileup'' of cosmic rays starting out at higher 
energies and crowding up in energy space at or below the predicted cutoff 
energy (Puget Stecker and Bredkamp 1976; Hill and Schramm 1985; Berezinsky and
Grigor'eva 1988; Stecker 1989; Stecker and Salamon 1999).
The existence and intensity of this predicted pileup depends critially on the
flatness and extent of the source spectrum, ({\it i.e.}, the number of cosmic 
rays starting out at higher energies), but if its existence is confirmed in the
future by more sensitive detectors, it would be evidence for the GZK effect.

Scully and Stecker (2000) have determined the GZK energy, defined as the
energy for a flux decrease of $1/e$, as a function of redshift. At high
redshifts, the target photon density increases by $(1+z)^3$ and both the 
photon and initial cosmic ray energies increase by $(1+z)$. The results
obtained by Scully and Stecker are shown in Figure 3.

\section{The Dog in the Night Time}

The lack of the expected ``GZK cutoff'' in the spectrum of ultrahigh energy
cosmic rays is a case of the dog that did nothing in the night 
time\footnote{In the {\it Adventure of Silver Blaze} by 
Arthur Conan Doyle, Sherlock Holmes noted the ``curious incident of the dog
in the night time''. The absence of action on the part of the dog was an 
important clue to solving the mystery.} It is an important clue, possibly 
eliminating some suggested astrophysical sources for these cosmic rays and
possibly pointing the way to new high energy physics.

\section{The Hunt for the Zevatrons}

\begin{figure}
\centerline{\psfig{figure=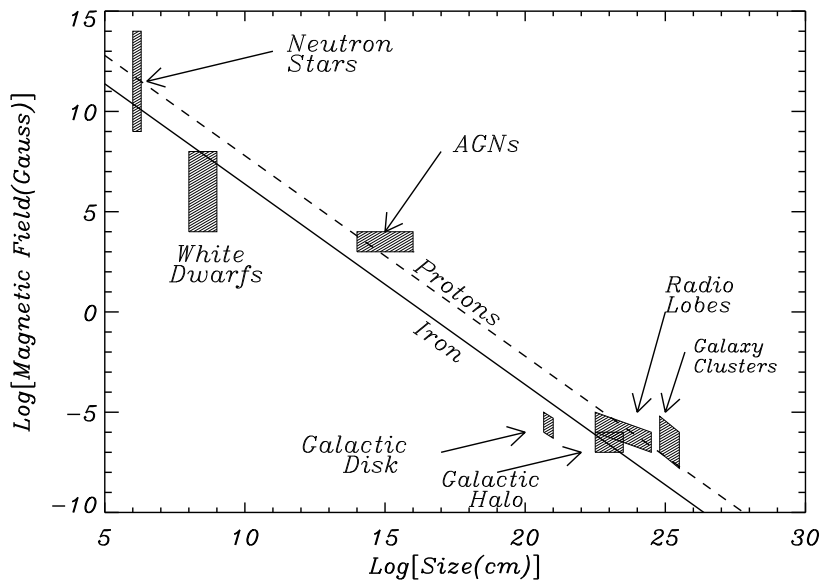,height=8cm}}

\caption{A ``Hillas Plot'' showing potential astrophysical zevatrons (from
Olinto (2000).}

\end{figure}

The lack of a GZK cutoff has led theorists to go on a hunt for nearby
``zevatrons'', {\it i.e.}, astrophysical sources which can accelerate 
particles to energies $\cal{O}$(1 ZeV = 10$^{21}$eV).

In most theoretical work in cosmic ray astrophysics, it is generally assumed 
that the diffusive shock acceleration process is the most likely mechanism for
accelerating particles to high energy (See, {\it e.g.}, Jones (2000) and
references therein). In this case, the maximum obtainable
energy is given by $E_{max}=keZuBL$, where $u \le c$ is the shock speed, 
$eZ$ is the charge of the particle baing accelerated, $B$ is the magnetic field
strength, $L$ is the size of the accelerating region and the numerical 
parameter $k = \cal{O}$$(1)$ (Drury 1994). Taking $k = 1$ and $u = c$, one 
finds

$$ E_{max} = 0.9Z(BL) $$

\noindent with $E$ in EeV, $B$ in $\mu$G and $R$ in kpc. This assumes that 
particles can be accelerated efficiently up until the moment when they can 
no longer be contained by the source, {\it i.e.} until their gyroradius 
becomes larger
than the size of the source. Hillas (1984) used this relation to construct
a plot of $B$ {\it vs.} $L$ for various candidate astrophysical objects. A
``Hillas plot'' of this kind, recently constructed by Olinto (2000), is shown 
in Figure 4.
 
Given the relationship between $E_{max}$ and $BL$ as shown in Figure 4, there 
are not too many astrophysical candidates for zevatrons. Of these, galactic
sources such as white dwarfs, neutron stars, pulsars, and magnetars can be 
ruled out because their galactic distribution would lead to anisotropies above
10 EeV which would be similar to those observed at lower energies by
Hayashida {\it et al} (1999), and this is not the case. Perhaps the most 
promising potential zevatrons are radio lobes of strong radio galaxies 
(Biermann and Strittmatter (1987). The trick is that such sources need to be 
found close enough to avoid the GZK cutoff ({\it e.g.}, Elbert and Sommers 
1995). Biermann (see these proceedings) has suggested
that the nearby radio galaxy M87 may be the source of the observed trans-GZK 
cosmic rays (see also Stecker 1968; Farrar and Piran 2000). Such an 
explanation would require one to invoke magnetic field
configurations capable of producing a quasi-isotropic distribution of 
$> 10^{20}$ eV protons, making this hypothesis questionable. However, if the 
primary particles are nuclei, it is easier to explain a radio galaxy
origin for the two highest energy events (Stecker and Salamon 1999; see 
section VII). 

It has also been suggested that since all large galaxies are suspected to
harbor supermassive black holes in their centers which may have once been
quasars, fed by accretion disks which are now used up, that nearby quasar
remnants may be the searched-for zevatrons (Boldt and Ghosh 1999; Boldt
and Lowenstein 2000). This scenario also has potential theoretical problems 
and needs to be explored further from a theoretical point-of-view.

Another proposed zevatron, the $\gamma$-ray burst, is discussed in the next 
section.

\section{Gamma-Ray Bursts: An Unlikely Suspect}

In 1995, it was suggested that cosmological $\gamma$-ray bursts (GRBs)
were the source of the highest energy cosmic rays (Waxman 1995; Vietri 1995).
It was suggested that if these objects emitted the same amount of energy in
ultrahigh energy ($\sim 10^{14}$ MeV) cosmic rays as in $\sim$ MeV photons, 
there would be enough energy input of these particles into intergalactic 
space to account for the observed flux. At that time, it was assumed that the 
GRBs were distributed uniformly, independent of redshift. 

In recent years, X-ray, optical, and radio afterglows of about a dozen
GRBs have been detected leading to the subsequent identification of the host
galaxies of these objects and consequently, their redshifts. The host galaxies of GRBs appear to be sites of active star
formation. The colors and morphological types of the host
galaxies are consistent with active star formation as
is the detection of Ly$\alpha$ and [OII] in several of these
galaxies. Further evidence 
suggests that bursts themselves are directly associated with star
forming regions within their host galaxies; their positions
correspond to regions having significant hydrogen column densities
with evidence of dust extinction.
It now seems more reasonable to assume
that a more appropriate redshift
distribution to take for GRBs is that of the average star
formation rate.

To date, some 14 GRBs afterglows have been detected with a subsequent
identification of their host galaxies.  At least 13 of the 14 are at moderate
to high redshifts with the highest one (GRB000131) lying at a redshift of
4.50 (Andersen, {\it et al.} 2000).\footnote{The origin of one burst, 
GRB980425, is somewhat controversial; a possible
X-ray source and an unusual nearby Type Ic
supernova have both been put forward as candidates.  Taking the
supernova identification gives an energy release which is 
orders of magnitude smaller 
than that for a typical cosmological GRB.}

A good argument in favor of strong redshift evolution for the frequency 
of occurrence of the GRBs has been made by Mao and Mo (1998), based on the
nature of the host galaxies. 
Other recent analyses have also favored a GRB redshift distribution 
which follows the strong redshift evolution of the star formation rate 
(Schmidt 1999; Fenimore and Ramirez-Ruiz 2000). 
If we thus assume a redshift distribution for the GRBs which follows the star
formation rate, being significantly higher at higher redshifts, 
GRBs fail by at least an order of magnitude to account
for the observed cosmic rays above 100 EeV (Stecker 2000). If one wishes to
account for the GRBs above 10 EeV, this hypothesis fails by two to three 
orders of magnitude (Scully and Stecker 2000). Even these numbers are most
likely to be too optimistic, since they are based on the questionable 
assumption of the same amount of GRB energy going into ultrahigh energy cosmic
rays as $\sim$ MeV photons.

Figure 5, from Scully and Stecker (2000), shows the {\it form} of the cosmic 
ray spectrum to be expected from sources with a uniform redshift distribution 
and sources which follow the star formation rate. The required normalization 
and spectral index determine the energy requirements of any cosmological 
sources which are invoked to explain the observations. Pileup effects and GZK 
cutoffs are evident in the theoretical curves in this figure. As can be seen 
in Figure 5, the present
data appear to be statistically consistent with either the presence or the 
absence of a pileup effect. Future data with much better statistics are
required to determine such a spectral structure.

\begin{figure}
\centerline{\psfig{figure=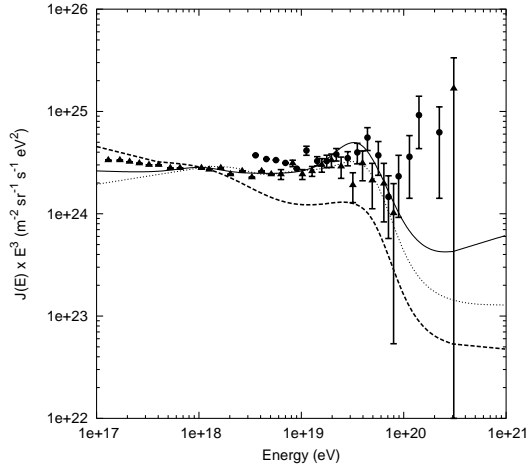,height=10cm}}

\caption{Predicted spectra for cosmic ray protons as compared with the data.
The middle curve and lowest curve assume an $E^{-2.75}$ source spectrum with
a uniform source distribution and one that follows the $z$ distribution of the
star formation rate respectively. The upper curve is for an $E^{-2.35}$ source
spectrum which requires an order of magnitude more energy input and exhibits
the ``pilup effect'' discussed in the text.}

\end{figure}

\section{Heavy Nuclei: A More Likely Suspect}

A more conservative hypothesis for explaining the trans-GZK events is that they
were produced by heavy nuclei. Stecker and Salamon (1999) have shown that the
energy loss time for nuclei starting out as Fe is longer than that for protons
for energies up to a total energy of 300 EeV (see Figure 6). 

Stanev {\it et al.} (1995) and Biermann (1998) have examined 
the arrival directions
of the highest energy events. They point out that
the $\sim 200$ EeV event is within 10$^\circ$ of the direction of the 
strong radio galaxy NGC 315.
NGC 315 lies at a distance of only $\sim$ 60 Mpc 
from us. For that distance, the results of Stecker and Salamon (1999) 
indicate that heavy nuclei would
have a cutoff energy of $\sim$ 130 EeV, which may be within the uncertainty in
the energy determination for this event. The $\sim$300 EeV event is within
12$^\circ$ of the direction of the 

\begin{figure}
\centerline{\psfig{figure=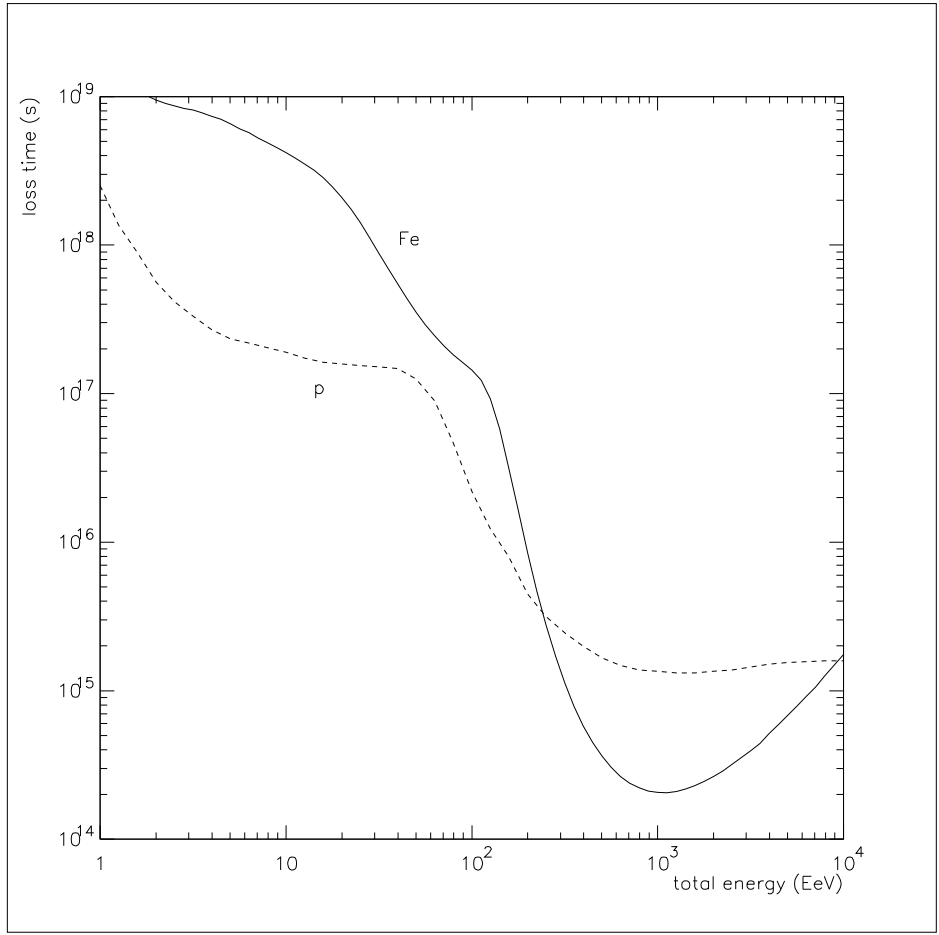,height=16cm}}

\caption{Mean energy loss times for protons (Stecker 1968; Puget, Stecker and
Bredekamp 1976) and nuclei originating as Fe (Stecker and Salamon 1999).}

\end{figure}

\noi strong radio galaxy 3C134. The distance to 3C134 
is unfortunately unknown because its location behind a dense molecular cloud 
in our own galaxy obscures the spectral lines required for a measurement of 
its redshift. It may be possible that {\it either} 
cosmic ray protons {\it or} heavy nuclei originated in 
these sources and produced the highest energy air shower events.

An interesting new clue that we may indeed be seeing heavier nuclei above the
proton-GZK cutoff comes from a very recent analysis of inclined air showers
above 10 EeV energy (Ave, {\it et al.} 2000). These new results favor
proton primaries below the p-GZK cutoff energy but they {\it appear to favor a 
heavier composition above the p-GZK cutoff energy}. It will be interesting to
see what future data from much more sensitive detectors will tell us.

\section{The ``Top'' Suspect (Spilling GUTs?)}

A way to avoid the problems with finding plausible astrophysical zevatrons is
to start at the top, {\it i.e.}, the energy scale associated with grand
unification, supersymmetric grand unification or its string theory equivalent.

  The modern scenario for the early history of the big bang takes account of 
the work of particle theorists to unify the forces of nature in the framework 
of Grand Unified Theories (GUTs). This concept extends the very successful 
work of Nobel Laureates Glashow, Weinberg,
and Salam in unifying the electromagnetic and weak nuclear forces of nature.  
As a consequence of this theory, the electromagnetic and weak forces would
have been unified at a higher temperature phase in the early history 
of the universe and then would have been broken into separate forces through
the mechanism of spontaneous symmetry breaking caused by vacuum fields, called
Higgs fields.

In GUTs, this same paradigm is used to infer that the electroweak force
becomes unified with the strong nuclear force at very high 
energies of $\sim 10^{24}$ eV 
which occurred only $\sim 10^{-35}$ seconds 
after the big bang. The forces then became separated owing to interactions
with the much heavier mass scale Higgs fields whose symmetry was broken
spontaneously. The supersymmetric GUTs (or SUSY GUTs) provide an explanation
for the vast difference between the two unification scales (known as the
``Hierarchy Problem'') and predict that the running coupling constants 
which describe the strength of the various forces become equal at the SUSY GUT
scale of $\sim 10^{24}$ eV.

The fossil remnants of this unification are predicted to be very heavy 
topological defects in the vacuum of
space caused by misalignments of the heavy Higgs fields in regions which
were causally disconnected in the early history of the universe. These are 
localized regions where extremely high densities of mass-energy are trapped. 
Such defects go by designations such as cosmic strings, monopoles, walls,
necklaces (strings bounded by monopoles), and textures, depending on their 
geometrical and topological properties.  Inside a topological defect, the 
vestiges of the early universe may be preserved to the present day.  
The general scenario for creating topological defects in the early universe 
was suggested by Kibble (1976).

Superheavy particles or topological structures arising at the GUT energy scale
$M \ge 10^{23}$ eV can decay or annihilate to produce ``X-particles'' (GUT 
scale Higgs particles, superheavy fermions, or leptoquark bosons of mass M.) 
In the case of strings
this could involve mechanisms such as intersecting and intercommuting
string segments and cusp evaporation. These X-particles will 
decay to produce QCD fragmentation jets at ultrahigh energies, so I will 
refer to them as ``fraggers''. QCD fraggers produce mainly pions with a 3 to
10\% admixture of 
baryons, so that generally one can expect them to produce at least an order of 
magnitude more ultrahigh energy $\gamma$-rays and neutrinos than protons. 
The same general scenario would hold for the decay of long-lived superheavy 
dark matter particles, which will also be fraggers. It has also been 
suggested that SUSY models which can
have an additional soft symmetry breaking scale at TeV energies (``flat SUSY
theories'') may help explain the observed $\gamma$-ray background flux at
energies $\sim$ 0.1 TeV (Bhattacharjee, Shafi and Stecker 1998). 

The number of variations and models for explaining the ultrahigh energy
cosmic rays based on the GUT or SUSY GUT scheme (which have come to be
called ``top-down'' models) 
has grown to be enormous and I will not attempt to list all of the
numerous citations involved. Fortunately, Bhattacharjee and Sigl (2000) have
recently published an extensive review with over 500 citations and I refer the
reader to this review for further details of ``top-down'' models and 
references. The important thing to note here
is that, if the implications of such models are borne out by future cosmic
ray data, they may provide our first real evidence for GUTs!

\subsection{``Z-bursts''}

It has been suggested that ultra-ultrahigh energy $\cal{O}$(10 ZeV) neutrinos
can produce ultrahigh energy $Z^0$ fraggers by interactions with  1.9K thermal
CBR neutrinos (Weiler 1982), resulting in ``Z-burst'' fragmentation 
jets, again mostly pions. This will occur at the resonance energy $E_{res} =
4[m_{\nu}({\rm eV})]^{-1}$ ZeV. A typical $Z$ boson will decay to produce
$\sim$2 nucleons, $\sim$20 $\gamma$-rays and $\sim$ 50 neutrinos, 2/3 of 
which are $\nu_{\mu}$'s. 

If the nucleons which are produced from Z-bursts originate within a few tens of
Mpc of us they can reach us, even though the original $\sim$ 10 ZeV 
neutrinos could have come from a much further distance. 
It has been suggested that this effect can be amplified if our galaxy has
a halo of neutrinos with a mass of tens of eV (Fargion, Mele and Salis 1999;
Weiler 1999). However, a neutrino mass large enough to be confined to a 
galaxy size neutrino halo would imply a hot dark matter cosmology which is 
inconsistent with simulations of galaxy formation and clustering 
({\it e.g.}, Ma and Bertschinger 
1994) and with angular fluctuations in the CBR. A mixed dark matter model with 
a lighter neutrino mass (Shafi and Stecker 1984) produces predicted 
CBR angular fluctuations (Schaefer, Shafi and Stecker 1989) which 
are consistent with the {\it Cosmic Background Explorer} data (Wright 1992). 
In such a model, neutrinos would have density fluctuations on the 
scale of superclusters, which would still allow for some amplification 
(Weiler 1999).

The basic general problem with the Z-burst explanation
for the trans-GZK events is that one needs to produce 10 ZeV neutrinos. If 
these are secondaries from pion production, this implies that the primary
protons which produce them must have energies of hundreds of ZeV! That is
why I have listed this possibility as a sub-section of ``top down'' models.
However, top-down models produce their own fraggers, making Z-bursts a 
secondary effect.

\section{Other New Physics Possibilities}

The GZK cutoff problem has stimulated theorists to look for possible solutions
involving new physics. Some of these involve (A) a large increase in the
neutrino-nucleon cross section at ultrahigh energies, (B) new particles, 
and (C) a small violation of Lorentz Invariance (LI).

\subsection{Increasing the Neutrino-Nucleon Cross Section at Ultrahigh 
Energies}

Since neutrinos can travel through the universe without interacting with the
2.7K CBR, it has been suggested that if the neutrino-nucleon cross section 
were to increase to hadronic values at ultrahigh energies, they could produce 
the giant air showers and account for the observations of showers above the
proton-GZK cutoff. Several suggestions have been made for processes that can
enhance the neutrino-nucleon cross section
at ultrahigh energies. These suggestions include 
composite models of neutrinos
(Domokos and Nussinov 1987; Domokos and Kovesi-Domokos 1988), scalar 
leptoquark  resonance channels (Robinett 1988) and the exchange of dual 
gluons (Bordes, {\it et al.} 
1998). Burdman, Halzen and Ghandi (1998) have ruled out
a fairly general class of these types of models, including those listed above,
by pointing out that in order to increase the neutrino-nucleon cross section
to hadronic values
at $\sim 10^{20}$ eV without violating unitarity bounds, the relevant scale 
of compositeness or particle exchange would have to be of the order of a 
GeV, and that such a scale is ruled out by accelerator experiments.

However, the interesting possibility exists for a large increase in the 
number of degrees of freedom above the electroweak scale in models of
TeV scale quantum gravity. It has been suggested that in such models,
$\sigma$($\nu$N) $\simeq [E_{\nu}/(10^{20}\rm eV)]$ mb (Nussinov and Schrock
1999; Jain, {\it et al.} 2000); see also Domokos and Kovesi-Domokos 1999). 
It should be noted that a cross section of
at least 10 mb would be necessary to approach obtaining consistency with
the air shower profile data.

\subsection{New Particles}

The suggestion has also been made that new neutral particles containing
a light gluino could be producing the trans-GZK events (Farrar 1996;
Cheung, Farrar and Kolb 1998). While the invocation of such new particles
is an intriguing idea, it seems unlikely that such particles of a few
proton masses would be produced in copious enough quantities in astrophysical
objects without being detected in terrestrial accelerators. Also there
are now strong constraints on gluinos (Alavi-Harati, {\it et al.} 1999).
One should note that while it is true that the GZK threshold for such 
particles would be higher than that for protons, 
such is also the case for the more prosaic heavy nuclei
(see section VII). In addition, such neutral particles cannot be accelerated 
directly, but must be produced as secondary particles, making the energetics
reqirements more difficult.

\subsection{Breaking Lorentz Invariance}  

With the idea of spontaneous symmetry breaking in particle physics came the
suggestion that Lorentz invariance (LI) might be weakly broken at high energies
(Sato and Tati 1972). Although no real quantum theory of gravity exists, it 
was suggested that LI might be broken as a consequence of such a theory
(Amelino-Camilia {\it et al.} 1998). A simpler formulation
for breaking LI by a small first order perturbation in the electromagnetic 
Lagrangian which leads to a renormalizable treatment has been given by
Coleman and Glashow (1999). Using this formalism, these authors have shown
than only a very tiny amount of LI symmetry breaking is required to avoid
the GZK effect by supressing photomeson interactions between
ultrahigh energy protons and the CBR. Of course, this would also eliminate
any ``pileup'' structure below the predicted GZK cutoff energy.

\section{Smoking Guns}

Future data which will be obtained with new detector arrays and satellites
(see next section) will give us more clues relating to the origin of the
trans-GZK events by distinguishing between the various hypotheses which have
been proposed.

A zevatron origin (``bottom-up'' scenario) will produce air-showers primarily
from primaries which are protons or heavier nuclei, with a much smaller 
number of neutrino induced showers. The neutrinos will be secondaries from 
the photomeson interactions with the 3K CBR photons (Stecker 1973; 1979; 
Hill and Schramm 1985; Stecker {\it et al} 1991).
In addition, zevatron events may cluster near the direction of the sources.

A ``top-down'' (GUT) origin mechanism will not produce any heavier nuclei and
will produce at least an order of magnitude more ultrahigh energy neutrinos
than protons. (For a discussion of ultrahigh energy neutrino astrophysics,
see Cline and Stecker 2000.) Thus, it will be important to look 
for the neutrino-induced air showers which are expected to originate 
much more deeply in the atmosphere than
proton-induced air showers and are therefore expected to be mostly
horizontal showers. Looking for these events can most easily be done with a
satellite array which scans the atmosphere from above (see next section).
The ``top down'' model also produces a large ratio of ultrahigh energy 
photons to protons, however, the mean free path of these photons against
pair-production interactions with extragalactic low frquency radio photons
from radio galaxies is only a few Mpc at most (Protheroe and Biermann 1996).
The subsequent electromagnetic cascade and synchrotoron emission of the
high energy electrons produced in the cascade dumps the energy of these 
particles into much lower energy photons (Wdowczyk, 
Tkaczyk and Wolfendale 1972; Stecker 1973).

Another distinguishing characteristic between bottom-up and top-down models
is that the latter will produce much harder spectra. If differential
cosmic ray spectra are parametrized to be of the form $F \propto E^{-\Gamma}$,
then for top-down models $\Gamma < 2$, whereas for bottom-up models
$\Gamma \ge 2$. Also, because of the hard source spectrum in the ``top-down'' 
models, they should exhibit both a GZK suppression and a pileup just before the
GZK energy.

If Lorentz invariance breaking is the explanation for the missing GZK effect,
the actual absence of photomeson interactions should result the absence of a
pileup effect as well.

\section{The New Detectives}

Of the ground-based ultra-high energy arrays, the AGASA array of particle
detectors in Japan continues to get ultrahigh energy cosmic ray data.
Its aperture is 200 km$^2$sr. 

The HiRes array is operating and will soon be publishing data. This array is 
an extension of the Fly's Eye which pioneered the technique of measuring the 
atmospheric fluorescence light in the near UV (300 - 400 nm range) that is
isotropically emitted by nitrogen molecules that are excited by the 
charged shower secondaries at the rate of $\sim$4 photons per meter per 
particle.
Its estimated aperture is 1000 km$^2$sr at 100 EeV with a 10\% duty cycle
(Sokolsky 1998) and it has already detected several events above the GZK 
cutoff energy (Abu-Zayyad, PhD thesis 2000).

The southern hemisphere Auger array is expected to be on line in the near 
future. This will be a hybrid array which will consist of 1600 particle
detector elements similar to those at Havera Park and three or four 
flourescence detectors. Its expected aperture will be 7000 km$^2$sr for the
ground array above 10 EeV and $\sim$ 10\% of this number for the hybrid array.

The Telescope Array will will consist of eight separate flourescence detecting
telescope stations separated by 30 km. Its expected aperture will be 8000 
km$^2$sr with an assumed 10\% duty cycle. 

The next big leap will be to go to a system of space-based detectors which
will look down on the Earth's atmosphere to detect the trails of nitrogen
flourescence light made by giant extensive air showers.
The Orbiting Wide-angle Light collectors (OWL) mission is being proposed to 
study such showers from satellite-based platforms in low 
Earth orbit (600 - 1200 km). OWL would observe extended air showers 
from space via the air fluorescence technique, thus determining the 
composition, energy, and arrival angle 
distributions of the primary particles in order to deduce their origin.  
Operating from space with a wide field-of-view instrument
dramatically increases the observed target volume, and consequently the 
detected air shower event 
rate, in comparison to ground based experiments. The OWL baseline 
configuration will yield event rates that are more than two orders 
of magnitude larger than currently operating ground-based experiments.  
The estimated aperture for a two satellite system is $3 \times 10^5$
km$^2$sr above a few tens of EeV assuming a 10\% duty cycle.

Figure 7 illustrates two OWL satellites obtaining stereoscopic views of an 
air shower produced by an ultra-high energy cosmic ray. 
With an approximate 10\% duty factor, OWL will be capable of
making accurate measurements of giant air shower events with high statistics. 
It is expected to be able to detect more than 1000 showers per year with 
$E \ge$ 100 EeV (assuming an extrapolation of the
cosmic ray spectrum based upon ground-based measurements). 

Closer in the future, the European Space Agency is 
studying the feasibility of placing such a light collecting detector on the 
International Space Station in order to develop the required technology to
observe the flourescent trails of giant extensive air showers, to make 
such observations,
and to serve as a pathfinder mission for a later free flyer. This experiment
has been dubbed the Extreme Universe Space Observatory (EUSO) (see paper
of Livio Scarsi, these proceedings, for more details).
Owing to the orbit parameters and constraints of the International
Space Station, the effective aperture for EUSO will not be as large 
as that of a free flyer mission.

A recent compendium of papers on observing giant air showers from space may be
found in Krizmanic, Ormes and Streitmatter (1998).

\begin{figure}
\centerline{\psfig{figure=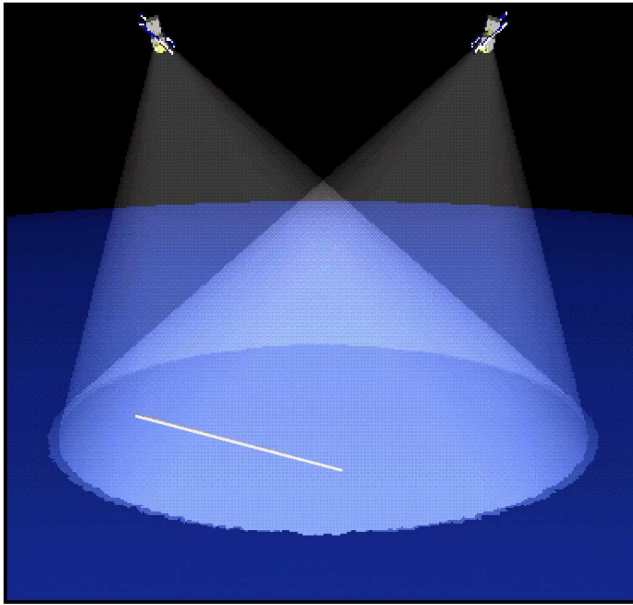,height=8cm}}

\caption{Two OWL satellites in low-Earth orbit observing the flourescent
track of a giant air shower. The shaded cones illustrate the field-of-view
for each satellite.}

\end{figure}

\section{The Mystery Remains}

With regard to the ultrahigh energy cosmic rays, we are, as Sherlock Holmes 
would have said, ``suffering from a plethora of surmise, conjecture, and
hypothesis''. Indeed, since ``The Adventure of Silver Blaze'', from which
I have now twice quoted, concerns a race horse, it seems appropriate 
language to say that many high energy theorists, given free rein, have
grabbed the bit between their teeth and bolted off to far pastures.

It seems fair to say that every solution which I have discussed in these
lectures, and more particularly, those which I have chosen not to discuss,
suffer from obvious problems or potential problems. Lets us then recall that
we could not understand how the sun shines until we had an understanding of
nuclear reactions. Perhaps future cosmic ray investigations will also lead
to one of those unexpected solutions in physics and astrophysics which will
result in truly new knowledge.




\acknowledgements
I would like to thank Tom Gaisser, Angela Olinto and Alan Watson for 
sending me files of their figures and for allowing me to use them in this 
paper. I would also like to thank John Krizmanic for his help with section XI 
and for supplying Figure 7.

{}

\end{document}